\documentclass[reprint,aps,prl]{revtex4-2}
\usepackage{amssymb,amsmath}
\usepackage{graphicx}
\pdfoutput=1

\usepackage{placeins}
\usepackage{bm}
\usepackage{epstopdf}
\usepackage{color}
\usepackage{xcolor}

\usepackage{subfigure}
\usepackage{parskip}
\usepackage[compact]{titlesec}
\usepackage{hyperref}

\usepackage{esvect}

\definecolor{mycolor}{RGB}{4, 55, 242}
\hypersetup{
	colorlinks=true,
	linkcolor=black,
	filecolor=blue,      
	urlcolor=mycolor,
	citecolor = black,breaklinks,backref}

\usepackage{cancel}

\usepackage{placeins}

\titlespacing{\section}{0pt}{1em}{0em}
\titlespacing{\subsection}{0pt}{1em}{0em}

\begin{document}
	
\title{Topological phase transition in fluctuating imaginary gauge fields }
\author{Bikashkali Midya}
\email{midya@iiserbpr.ac.in} 
\affiliation{Indian Institute of Science Education and Research Berhampur, Odisha 760003, India}

\begin{abstract}
	We investigate the exact solvability and point-gap topological phase transitions in non-Hermitian lattice models. These models incorporate site-dependent nonreciprocal hoppings $J e^{\pm g_n}$, facilitated by a spatially fluctuating imaginary gauge field $ig_n \hat{x}$ that disrupts translational symmetry. By employing suitable imaginary gauge transformations, it is revealed that a lattice characterized by any given $g_n$ is spectrally equivalent to a lattice devoid of fields, under open boundary conditions. Furthermore, a system with closed boundaries can be simplified to a spectrally equivalent lattice featuring a uniform mean field $i\bar{g}\hat{x}$. This framework offers a comprehensive method for analytically predicting spectral topological invariance and associated boundary localization phenomena for bond-disordered nonperiodic lattices. These predictions are made by analyzing gauge-transformed isospectral periodic lattices. Notably, for a lattice with quasiperiodic $g_n= \ln |\lambda \cos 2\pi \alpha n|$ and an irrational $\alpha$, a previously unknown topological phase transition is unveiled. It is observed that the topological spectral index $W$ assumes values of $-N$ or $+N$, leading to all $N$ open-boundary eigenstates localizing either at the right or left edge, solely dependent on the strength of the gauge field, where $\lambda<2$ or $\lambda>2$. A phase transition is identified at the critical point $\lambda\approx2$, at which all eigenstates undergo delocalization. { The theory has been shown to be relevant for long-range hopping models and for higher dimensions.}
\end{abstract}
\maketitle

Recently, there has been a growing interest in the physics of imaginary gauge field (IGF) within non-Hermitian condensed matter systems \cite{Longhi2018,Heusen2021,Rivero2022,Wong2021,Cheng2024,Descheemaeker2017,Pang2024,Gao2023,Faugno2022,Midya2018,Peng2015,Gao2024}. When a particle is subjected to a uniform IGF, denoted as $ig\hat{x}$, it acquires an imaginary Peierl's phase of $e^{-g}$ (or $e^{+g}$) as it tunnels to the nearest-neighbor site to the right (or left) within a lattice. This asymmetrical interaction enables directional  transport \cite{Longhi2015,Wanjura2020,Ramos2021,Midya2022,Wang2022} and boundary localization of all eigenstates, exemplifying the non-Hermitian skin effect (NHSE) \cite{Yao2018,Song2019,Lee2019,Haga2021}. This phenomenon is distinctively preserved due to the topological protection of the NHSE, stemming from the non-trivial point-gap topology of the bulk energy band \cite{Okuma2020,Borgnia2020,Gong2018}; however, the effect vanishes at the topological phase transition for zero IGF. The NHSE has been experimentally observed in various fields such as photonics \cite{Weidemann2020}, metamaterials \cite{Brandenbourger2019}, ultracold atoms \cite{Liang2022,Xiao2020}, quantum Hall devices \cite{Ochkan2024}, and reconfigurable laser systems \cite{Gao2023}.

 Much of the current understanding of IGF-induced topological NHSE relies on periodic structures \cite[and references therein]{Zhang2022}. Describing band topology and associated eigenstate localization in a lattice with spatially varying IGF, which hinders translational symmetry, poses a challenge. Traditional tools like the Fourier transform, Bloch’s theorem, and the Brillouin zone do not directly apply in such cases. Is there a method to alleviate these challenging circumstances?  Alternative strategies include computing  ``real space" topological invariants  of spatially disordered lattices~\cite{Hughes2021,Wang2024}, as well as the analysis of  Lyapunov exponent of a temporally fluctuating system \cite{Longhi2020}. The absence of a theoretical framework that can analytically elucidate the hidden ``reciprocal-space" topological phases and foresee the emergence of real-space NHSE in quasiperiodic and nonperiodic systems is a noteworthy limitation. 
 
 {\it IGF induced nearest-neighbor Hamiltonian.} Here, we present a theoretical framework for the analytical prediction of spectral topological invariance and associated localization phenomena, even in lattices with broken transnational symmetry due to nonuniform IGFs. To achieve this, we explore topological phases in  a Hatano-Nelson lattice \cite{Hatano1996,Nelson1998} with an arbitrary IGF $ig_n\hat{x}$, where $g_n\in\mathbf{R}$. A corresponding nonperiodic Hamiltonian  is given by
\begin{equation}
H\{g_n\}=J \sum_{n=1}^{N-1} \left(e^{g_n} c_n^+ c_{n+1}^- + e^{-g_n} c_{n+1}^+ c_n^-\right) + H_{\mathrm B},\label{Hamil-1}
\end{equation}
 where $Je^{g_n}$ and $J e^{-g_n}$ are IGF induced left- and right-nearest-neighbor hopping amplitudes, respectively. $N$ is the total number of sites.  $c_n^+$  $(c_n^-)$ are spinless particle creation  (annihilation) operators at site $n$. The boundary condition is specified as follows: for an open boundary lattice, it is denoted by $H_\mathrm{B}=0$, whereas for a closed boundary lattice, it is given by $H_\mathrm{B}=J(e^{g_{_N}} c_N^+c_1^- + e^{-g_{_N}} c_1^+ c_N^-)$.  Note that the non-Hermitian Hamiltonian mentioned above appears in semi-classical description of open quantum systems, specifically in cases where the effect of the quantum jump can be disregarded~\cite{Nori2019}. In this approximation, the system density matrix evolves according to the reduced master equation: $\dot{\rho}\simeq-i(H\rho-\rho H^\dag)$, where the effective-Hamiltonian 
$
H=H_0-i\sum_n \ell_n^+ \ell_n^-, \label{heff}
$
 contains the system Hamiltonian $H_0=J\sum_n (c_n^+ c_{n+1}^-+c_{n+1}^+c_n^{-})$ generating the coherent dynamics and the nonlocal Lindblad jump operators  $\ell_n^\pm=\sqrt{\gamma_n}(c_n^\pm\mp ic_{n+1}^\pm)$ characterizing the dissipative system-environment interaction with non-uniform couplings $\gamma_n$. Apart from an irrelevant constant imaginary shift of energy, the effective Hamiltonian reduces to \eqref{Hamil-1} provided one identifies the gauge field $g_n=\frac{1}{2}\ln \left|\frac{J+\gamma_n}{J-\gamma_n}\right|.$

{\it Open boundary lattice and NHSE.}   The spectral and localization properties of an open Hatano-Nelson chain are described using an imaginary gauge transformation (IGT)
\begin{equation}
c_n^{ \pm} \mapsto \tilde{c}^\pm_n = e^{\mp G_n}~ {c}_n^{ \pm}, \label{IGT1}
\end{equation}
where  $G_1=0, \quad 	G_n=\sum_{j=1}^{n-1} g_j. $
The above transformation leaves the commutator and anticommutator invariant: $[c_n^-,c_n^+]=[\tilde{c}_n^-,\tilde{c}_n^+]$ and $\{c_n^-,c_n^+\}=\{\tilde{c}_n^-,\tilde{c}_n^+\}$, respectively. $G_n$ has the following physical interpretation: it is the total imaginary phase accumulated by the particle while hopping  in the direction $1\leftarrow n$, whereas $-G_n$ corresponds to the phase acquired during the $1\rightarrow n$ hopping.  Within the framework of the IGT (4), the gauge field is entirely eliminated, leading to a transformation of the Hamiltonian $H$ into its isospectral equivalent. This transformed Hamiltonian corresponds to a uniformly coupled lattice, as described by the  Hermitian form
 \begin{equation}
 	H\{g_n\}\mapsto\widetilde{H}\{0\}=J \sum_{n=1}^{N-1} \left( \tilde{c}_n^+ \tilde{c}_{n+1}^- +  \tilde{c}_{n+1}^+ \tilde{c}_n^-\right), \label{Htilde}
 \end{equation}
 with real spectrum $ E_j= 2 J \cos\frac{\pi j}{N+1}, \quad j=1,2,\cdots, N.$

 When expressed in matrix forms, the above gauge transformation can be expressed in terms of a similarity transformation: $\widetilde{H}=
 \mathcal{G} H \mathcal{G}^{-1}$, where $\mathcal{G}=\operatorname{diag}(e^{G_1},\cdots,e^{G_N})\in \mathrm{GL}_N(\mathbf{R})$ is nonunitary.	Note that the Hermiticity of the Hamiltonian $\widetilde{H} = \widetilde{H}^\dag$ is ensured provided $(\tilde{c}_n^+)^\dag=\tilde{c}_n^-$.  This is  contrary to the non-reciprocal system described by $H$ in Eq.~\eqref{Hamil-1}, where the operators  $(c_n^+)^\dag \ne c_n^-$, but are pseudo-Hermitian conjugate to each other, i.e., 
$(c_n^+)^\dag = e^{2G_n} c_n^-$ follows from the IGT~\eqref{IGT1}.  This reveals the inherent pseudo-Hermitian character \cite{Mostafazadeh2010}  $H^\dag=\eta H \eta^{-1}$, where $\eta=\mathcal{G}^2$ is a positive definite operator that  guarantees the entirely real spectrum of $H$. 

The eigenstates $|\Psi(E_j)\rangle=\sum_n\psi_n(E_j) c_n^+|0\rangle$ of the Schr\"odinger equation $H|\Psi_j\rangle=E_j|\Psi_j\rangle$ can be conveniently derived from those of $\widetilde{H}$ using the relation $|\Psi_j\rangle = \mathcal{G}^{-1}|\widetilde{\Psi_j}\rangle$, where the probability amplitude at site $n$ is given by  
 \begin{equation}
 	\psi_n(E_j) = \sqrt{\frac{2}{N+1}}~e^{-G_n} \sin \frac{n j\pi}{N+1}.
 \end{equation}
 These states are nonorthogonal: $\sum_n\psi_n(E_j) \psi_n(E_{j'})\ne \delta_{jj'}$. Instead, they exhibit pseudo-orthogonality with respect to $\eta$:
$\sum_n \eta_{nn}~\psi_n(E_j) \psi_n(E_{j'}) =\delta_{jj'}$ [which is the biorthogonality relation with respect to right and left eigenstates $|\Psi(E_j)\rangle$ and $\eta|\Psi(E_{j'})\rangle$, respectively]. It is worth noting that all energy eigenstates satisfy the bound
\begin{equation}
	|\psi_n(E_j)|\le e^{-G_n}, \quad \text{for all} ~j, n \label{psin}
\end{equation}
indicating the possibility for exponential confinement of all eigenstates. Generally, the value of $e^{-G_n}$ peaks when the phase $G_n$ is minimized. Consequently, NHSE can manifest in the scenario of a monotonically changing $G_n$ that attains a global minimum at one of the lattice boundaries. The precision of localization diminishes when the monotonicity criterion is weakened. Some specific cases warrant consideration. First, when $G_n$ exhibits monotonically increasing behavior (e.g., when $g_n\ge0$), the skin effect manifests at the left boundary. Secondly, when $G_n$ monotonically decreases (e.g., when $g_n\le0$), the skin effect arises at the right boundary of a lattice. In both instances, it further holds that $|\psi_n(E_j)|\lesssim e^{-n \bar{g}}$, with the localization length approximating $|\bar{g}|^{-1}$ derived as the reciprocal of the mean gauge field  per site:
 \begin{equation}
  \bar{g}=\frac{1}{N}\sum\limits_{n=1}^N {g_n}.
 \end{equation}
 Thus boundary localization occurs for any arbitrary $g_n$ under the conditions that (i) $G_n$ is monotonic (at least weakly) and (ii) the associated value of $\bar{g}$ is non-zero. Furthermore, the localization at either the left or right boundary is contingent on whether $\bar{g}$ is greater than or less than zero; localization is sharper for higher value $\bar{g}$. The appearance of NHSE is not guaranteed in the event where $\bar{g}$ equals zero.  In this context, $\bar{g}$ serves as an ``order parameter" that quantifies the degree of ``global" non-reciprocity in the system. A null value for $\bar{g}$ indicates that the system is effectively reciprocal (despite being nonreciprocal at a local level) and exhibits characteristics akin to a Hermitian system with all states being delocalized.

\begin{figure*}[ht!]
	\centering{\includegraphics[width=0.95\textwidth]{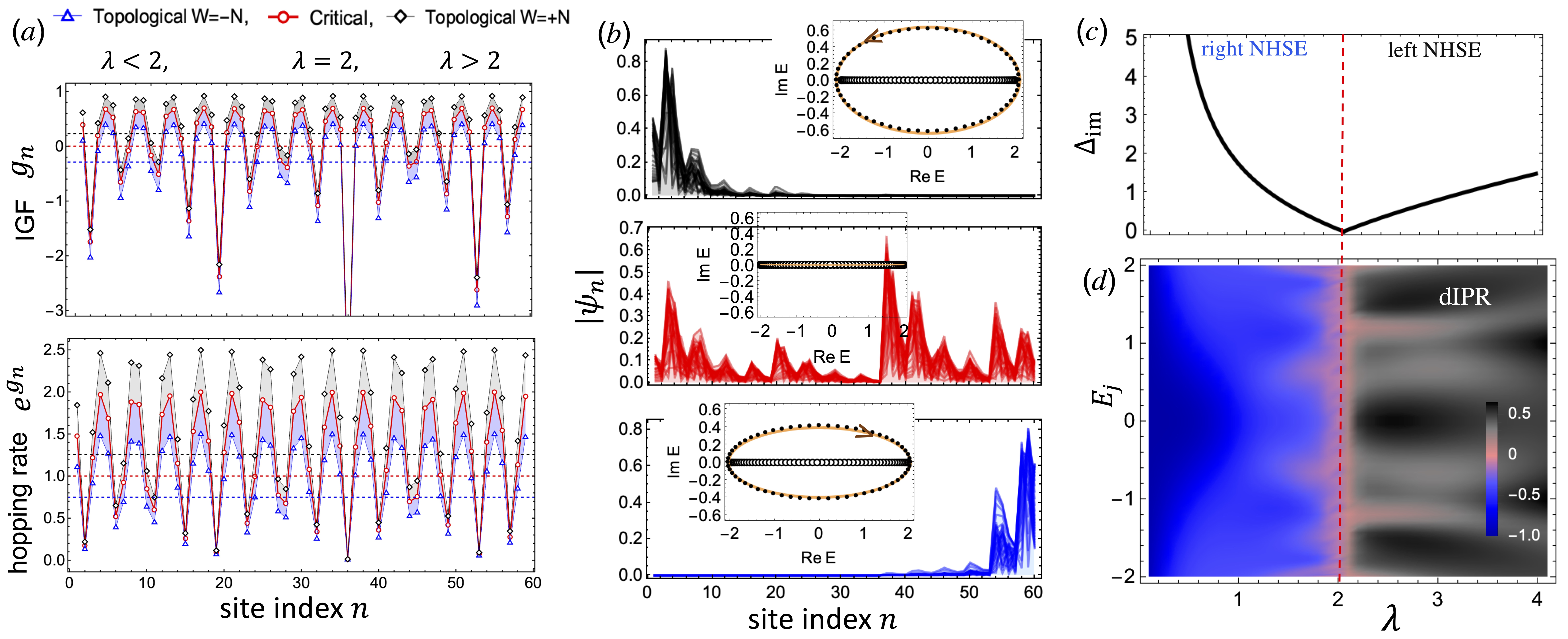}}
	\caption{Topological phase transition and NHSE in a quasi-periodic IGF. (a) The upper panel illustrates field distribution $g_n$, corresponding to $\lambda=1.5$ (blue, triangle), $\lambda=2$ (red, circle), and $\lambda=2.5$ (black, diamond). Additionally, the lower panel presents position-dependent left-hopping rates $J e^{g_n}$. The mean-field value $\bar{g}(\lambda)$ is depicted by dashed lines in the upper panel, along with the uniform hopping $Je^{\bar{g}}$ in the lower panel where $\bar{g}(1.5)=-0.3, \bar{g}(2)\sim0, \bar{g}(2.5)=0.2$.  (b) Eigenstates of an open Hatano-Nelson chain exhibit NHSE at both the left edge (upper panel for $\lambda=2.5$) and right edge (lower panel for $\lambda=1.5$), while boundary localization is absent near the critical point ($\lambda=2.06$, middle panel). The open and closed boundary spectra of $H$ are depicted by open circles and black dots in the insets, respectively. {The corresponding mean-field spectra (orange line) are presented for comparison.} (c)  The imaginary gap of the closed boundary spectrum against varying values of $\lambda$ is provided. (d) Illustrates the dIPR for all open-boundary eigenstates as $\lambda$ varies around the critical point demonstrating a phase transition characterized by the crossover from negative to positive values of dIPR. Here, $J=1$ and $N=60$. Quantities plotted here are in dimensionless units. } \label{fig-1}
\end{figure*}

{\it Closed boundary lattice and topological phases.} In order to uncover the topological origin of the NHSE mentioned above, we will now examine a closed Hatano-Nelson chain. In this scenario, the initial nonperiodic lattice containing randomly distributed IGF  $ig_n\hat{x}$ can be drastically simplified to a periodic lattice with a spatially uniform field $i\bar{g}\hat{x}$ and associated hopping rates $Je^{\pm \bar{g} }$. The transformed mean-field Hamiltonian is described by $H\{g_n\}\mapsto\widetilde{\widetilde{H}}\{\bar{g}\}$, where 
\begin{equation} \widetilde{\widetilde{H}}\{\bar{g}\}=J\sum\limits_{n=1}^N \left(e^{\bar{g}} \tilde{\tilde{c}}_n^+ \tilde{\tilde{c}}_{n+1}^- + e^{-\bar{g}} \tilde{\tilde{c}}_{n+1}^- \tilde{\tilde{c}}_n^+	\right). \label{HPBC}
\end{equation}
Here, $N+1\equiv 1$ is assumed to satisfy the closed boundary condition. In order to acquire $\widetilde{\widetilde{H}}$, we shall now use the subsequent IGT [the IGT defined in  Eq.~\eqref{IGT1} is deemed ineffective for a lattice that is subjected to a closed boundary condition]
\begin{equation}
	c^\pm_{n}\mapsto \tilde{\tilde{c}}_n^\pm= e^{\mp G_n \pm (n-1)\bar{g}} {c}_n^\pm, \label{IGT2}
\end{equation}
which satisfies $c^\pm_{N+p}\equiv c^\pm_{p}$ and $ \tilde{\tilde{c}}_{N+p}^\pm\equiv  \tilde{\tilde{c}}_p^\pm$ if $g_{_N+p}$ is identified with $g_p$ in a closed lattice for integer $p=1,2,\cdots$. In the presence of a constant IGF $g_n = g$, this transformation loses significance as the exponent disappears. However, it is pivotal in comprehending non-constant IGFs. The property of being translation invariant in $\widetilde{\widetilde{H}}$ allows for the convenient application of the Fourier transform. In the Fourier representation  $\tilde{\tilde{c}}_n^\pm= \int_0^{2\pi} dk ~\frac{e^{\mp i k n}}{\sqrt{2\pi}}  ~ \tilde{\tilde{c}}_k^{\pm}$, one can readily derive
\begin{equation}
\widetilde{\widetilde{H}}=\int_0^{2\pi} dk ~E(k)~ \tilde{\tilde{c}}_k^{+} \tilde{\tilde{c}}_k^{-},
\end{equation}
where the spectrum is complex and takes the form $
E(k)= 2 J \cos \left(k-i \bar{g}\right)$.  $E(k)$ traces a counterclockwise (clockwise) elliptical loop
\begin{equation}
	\frac{(\mathrm{Re} E)^2}{\cosh^2 \bar{g} } + \frac{(\mathrm{Im}E)^2}{\sinh^2 \bar{g}} = 4 J^2,
\end{equation}
in the complex energy plane as a function of $k$ in the Brillouin zone when $\bar{g}$ is positive (negative). The imaginary point-gap is given by $\Delta_{\mathrm{im}}=|2J\sinh \bar{g}|$ and corresponds to the minor axis of the ellipse. This gap persists as long as $\bar{g}$ is not zero. As a result, the topological characteristic of the point gap in the system is derived from the winding number of the loop of $E(k)$ with respect to all interior open boundary spectral points $E_j$~\cite{Gong2018}
\begin{equation}
	w(E_j)=\frac{1}{2\pi i}\oint_{0}^{2\pi} dk ~ \partial_k \ln[E(k)-E_j].
\end{equation}
Total winding number is given by $W=\sum_{j=1}^N w(E_j)$. It turns out that $W$ quantizes to $+N$ or $-N$ depending on $\bar{g} >0$ or $<0$. As a result, there are $N$ number of localized edge states either on the left or right boundary of the open lattice. The topology is undefined when imaginary-gap closes ($\Delta_{\mathrm{im}}=0$) for $\bar{g}=0$, which indicates the onset of a topological phase transition.

\begin{figure}[t!]
	\centering{
		\includegraphics[width=0.4\textwidth]{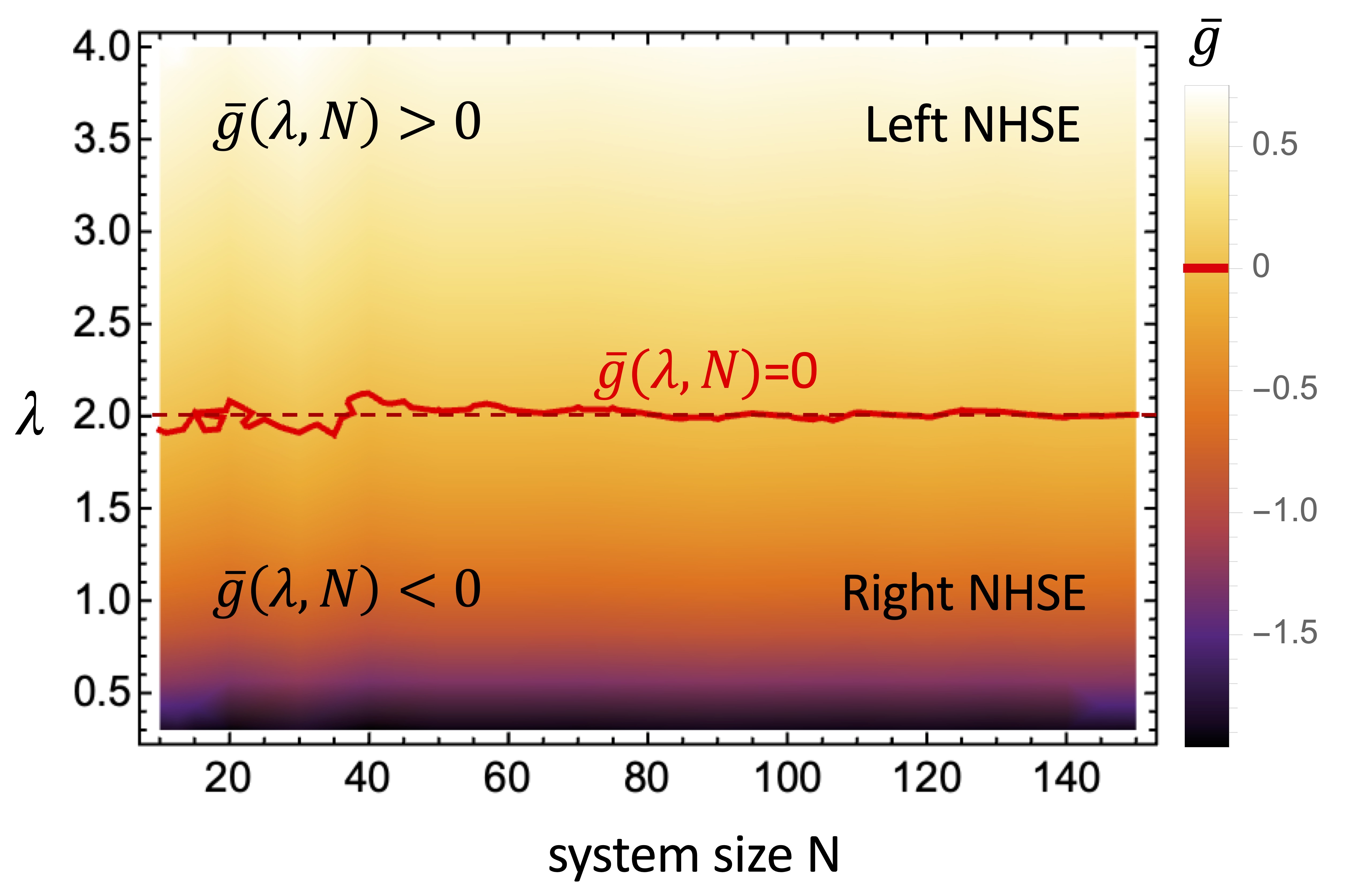} }
	\caption{The mean-field $\bar{g}$ (without approximation, and in dimensionless unit) is depicted as a function of field strength $\lambda$ and system's size $N$ in the context of the IGF \eqref{IGF}. The red line represents critical points at which $\bar{g}=0$, separating between left- and right-NHSE regimes.} \label{fig-2}
\end{figure}

{\it Application to a quasiperiodic IGF model.} In the special case, when IGF $g_n=g, \forall n$, we recover the standard NHSE of a uniform Hatano-Nelson lattice \cite{Zhang2022}. To demonstrate the effectiveness of the theory, we now consider a nontrivial example of a lattice with quasiperiodic IGF defined by 
\begin{equation}
	g_n = \ln \lambda |\cos 2\pi \alpha n|,\label{IGF}
	\end{equation}
with irrational $\alpha = (\sqrt{5}-1)/2$.  $\lambda>0$ characterizes the uniform field strength equal to $\ln \lambda$ and local fluctuations are given by $\ln|\cos2\pi\alpha n|$ [see Fig.~1a]. The associated mean-field is computed for a range of values of $\lambda$ and $N$ (see Fig.~\ref{fig-2}). For a sufficiently large $N$ ($\gtrsim50$),  one obtains  \cite{suppl}
\begin{equation}
 \bar{g}(\lambda)  = \ln \lambda \left(\prod\limits_{n=1}^{N} |\cos 2\pi \alpha n| \right)^\frac{1}{N} \simeq \ln \frac{\lambda}{2}.\label{mean-field}
	\end{equation}
It is positive for values of $\lambda$ greater than 2 and negative for values less than 2. As per the theory discussed above, the occurrence of the topological NHSE is expected at the end of the lattice, determined solely by the field intensity $\lambda$ being either less than $2$ (for the right edge) or greater than $2$ (for the left edge). When $\lambda$ equals the critical value $2$, all eigenstates lose their localization due to the mean field being identically zero. This occurrence is also to be anticipated, as the imaginary band gap becomes zero at $\lambda=2$ based on the expression 
\begin{equation}
	\Delta_{\mathrm{im}}(\lambda)\simeq \left|2J \sinh \left(\ln \frac{\lambda}{2}\right)\right|.
	\end{equation}
	At the critical $\lambda$, the lattice becomes globally reciprocal. This is proven by the fact that the geometric mean of all leftward hoppings given by $Je^{\bar{g}}$ matches that of rightward hoppings $Je^{-\bar{g}}$, where $\bar{g}=0$ at $\lambda\sim2$. Our main findings presented in Figure 1 illustrate the manifestations of left- and right- skin effect and the phenomenon of delocalization, induced by the respective distributions of  IGFs ($g_n$), for a few values of $\lambda$ in a lattice comprising $60$ sites.

\begin{figure}[t!]
	\centering{
		\includegraphics[width=0.45\textwidth]{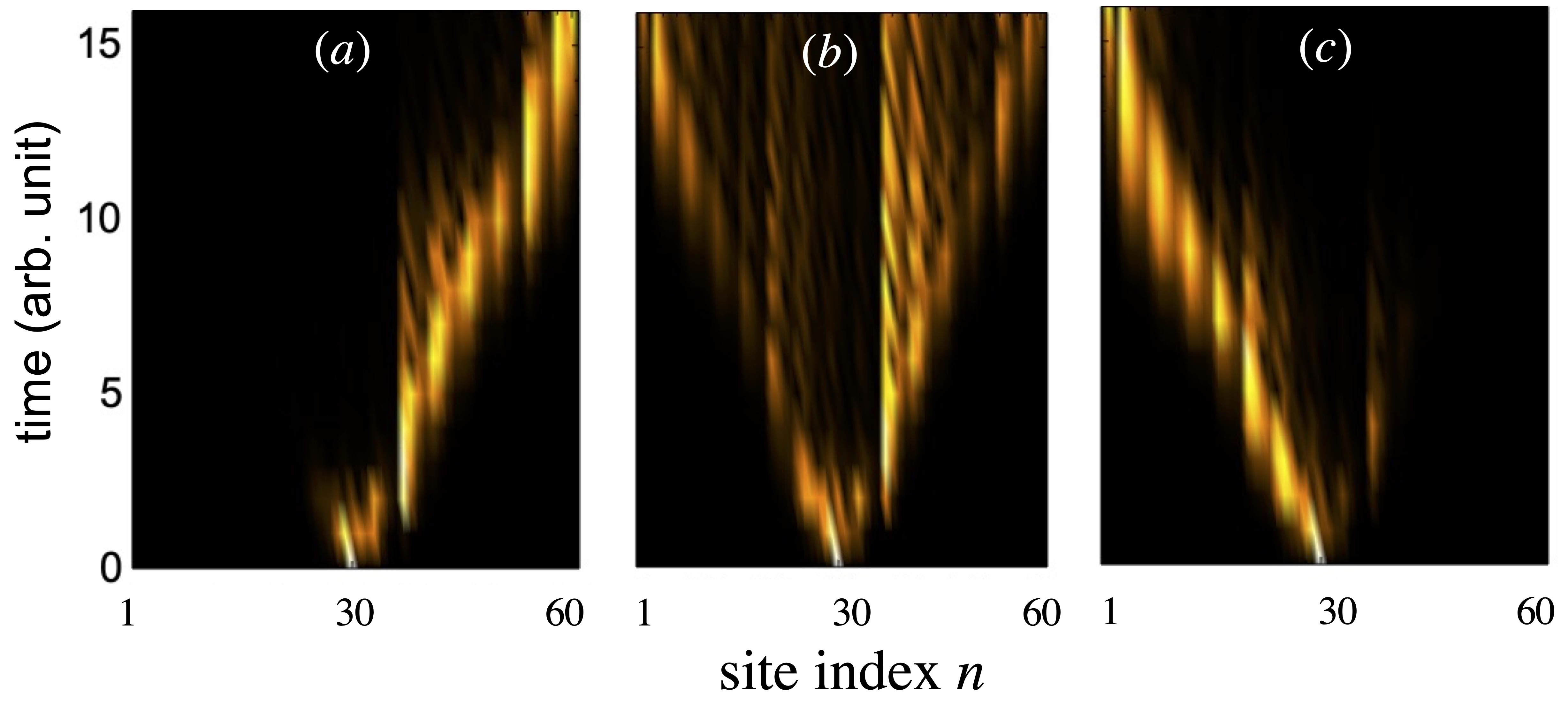} }
	\caption{ Temporal evolution of an initial excitation at the center of a lattice is shown for three distinct scenarios: (a) when $\lambda=1.5$, (b) when $\lambda=2.06$, and (c) when $\lambda=2.5$. Notably, the phenomena of NHSE-assisted directed transport and probability funneling are observed in cases (a) and (c). Conversely, in scenario (b), where NHSE is absent, the evolution displays dispersive characteristics. Here time is in arbitrary unit. } \label{fig-3}
\end{figure}

The degree of localization of the $j$th eigenstate $|\Psi_j\rangle$, of an open boundary lattice, is measured by the inverse of the participation ratio, denoted as $\mathrm{IPR}(|\Psi_j\rangle)=\sum_n|\psi_n(E_j)|^4/\left(\sum|\psi_n(E_j)|^2\right)^2$. A value close to $1$ for the IPR indicates localized states, while a value close to $0$ signifies extended states. The localization center, either at the left or right side of a lattice, can be determined  by the directional IPR defined as~\cite{Zeng2022}: $\mathrm{dIPR}(|\Psi_j\rangle)=\sum\limits_{n=1}^N \mathrm{sgn}\left[(N/2-n)|\psi_n(E_j)|\right]\mathrm{IPR}(|\Psi_j\rangle)$, here $\mathrm{sgn}[x]$ is the signum function.  The dIPR is positive when $|\Psi_j\rangle$ is localized at the left boundary, while it becomes negative when the eigenstate is localized at the right boundary.  In Fig.~1(d), we have depicted the dIPR for all eigenstates as the parameter $\lambda$ varies across the critical value. We also examine the evolution over time of a localized excitation $|\Phi_0\rangle=c^+_{N/2}|0\rangle$, initiated at $t=0$. Here, $|0\rangle$ symbolizes an unoccupied lattice. The wave function at time $t$ is expressed as $|\Phi(t)\rangle=e^{-i H t}|\Phi_0\rangle/||e^{-iHt}|\Phi_0\rangle||$. The time-dependent single-particle probability density at site $n$ is determined by $P(n,t)=\langle \Phi(t)| c_n^+c_n^-|\Phi(t)\rangle$. Illustrated in Fig. 3, the initial excitation at the center of a lattice demonstrates a directional propagation to the right (left) for $\lambda<<2$ ($\lambda>>2$), subsequently accumulating at the boundary, exemplifying funneling of the probability wave. In contrast, a distribution of the wave extends across the lattice, leading to the suppression of the funneling effect as $\lambda\sim2$ is approached.

{\it Generalization to long-range hopping Hamiltonian.} The theory can be extended beyond nearest-neighbor interaction by considering a long-range hopping Hamiltonian under IGF framework \cite{suppl}:
\begin{equation}
	H= \sum_{p=1}^{P}\sum_{n=1}^{N-p} J^{(p)}\left(e^{-i\theta_n^{(p)}} c^+_n c^-_{n+p} + e^{i\theta_n^{(p)}} c^+_{n+p} c^-_{n}\right) +H_B^{(p)}.\label{long-range}
\end{equation}
Here, $J^{(p)}$ represents the real hopping amplitude associated with the $p$th nearest-neighbor tunneling and $\theta_n^{(p)}=i(g_n+g_{n+1}+\cdots+g_{n+p-1})$ denotes the imaginary Peierl's phase. The boundary term for a closed lattice (where $N+p\equiv p$) is given by
\begin{equation}
	H_B^{(p)}=\sum_{n=N-p+1}^{N}{\hspace{-.5em}}J^{(p)}(e^{-i\theta_n^{(p)}} c_n^+c_{n+p}^- + e^{i\theta_n^{(p)}} c_{n+p}^+c_{n}^-).
\end{equation}
The implementation of the IGT \eqref{IGT1}  annihilates the imaginary phase and non-Hermiticity  from the above Hamiltonian \eqref{long-range} featuring open boundaries, resulting in a Hermitian Hamiltonian characterized by long-range hoppings $J^{(p)}$. As the latter exhibits fully delocalized states, it follows that the states corresponding to $H$ become localized as a consequence of IGT. On the other hand, a closed boundary lattice undergoes a transformation into the following long-range mean-field Hamiltonian~\cite{suppl}:
\begin{equation}
	\tilde{\tilde{H}}= \sum_{p=1}^{P}\sum_{n=1}^{N} J^{(p)}\left(e^{p\bar{g}}~\tilde{\tilde{c}}^+_n \tilde{\tilde{c}}^-_{n+p} + e^{-p \bar{g}}~ \tilde{\tilde{c}}^+_{n+p} \tilde{\tilde{c}}^-_{n}\right).
\end{equation}
as a result of IGT \eqref{IGT2}. The corresponding energy band can be expressed as: $E(k)=\sum\limits_{p=1}^P 2 J^{(p)} \cos p(k-i\bar{g}),$ which notably exhibits higher-order spectral winding in the complex energy plane for non-zero values of $\bar{g}$ \cite{Wang2021,suppl}. In \cite{suppl}, numerical demonstrations of open boundary NHSE and the related closed boundary spectral topology are presented for a quasiperiodic IGF lattice featuring nearest and next-nearest-neighbor hoppings.

{\it Generalization to complex gauge field.}  Lattices subjected to gauge fields represented as $g_n=g_n'+ig_n''$ can also be analyzed using the theory mentioned above. In this scenario, the NHSE and underlying topology are solely dependent on the real components of $g_n$, $G_n$ and $\bar{g}$, as the non-Hermiticity and non-reciprocity in a system are unaffected by their respective imaginary components.

{\it Generalization to two-dimensions (2D).} A  rectangular lattice under 2D IGF $(ig_n\hat{x}+ih_m\hat{y})$ is governed by the Hamiltonian $H_{2D}=\sum\limits_{nm}J_x( e^{g_n}c_{n,m}^+c_{n+1,m}^-+e^{-g_n}c_{n+1,m}^+c_{n,m}^-)+J_y (e^{h_m}c_{n,m}^+c_{n,m+1}^-+e^{-h_m}c_{n,m+1}^+c_{n,m}^-)$, subject to appropriate boundary conditions. As elaborated in the Supplemental Material \cite{suppl}, on a torus geometry, $H_{2D}\{g_n,h_m\}$ can be reduced to a spatially uniform Hamiltonian $\widetilde{\widetilde{H}}\{\bar{g},\bar{h}\}$ by a 2D IGT, where  $\bar{g}$ and $\bar{h}$ denote mean-fields along $x$ and $y$ directions, respectively.  The associated ``weak" spectral topology ${\bf{w}}=(w^x,w^y)$ \cite{Kawabata2020,Gao2023}, derived from one-dimensional winding numbers for the $x$ and $y$ directions, characterizes corner NHSE occurrences within an open lattice when non-zero mean-fields are present in both directions.

{\it Experimental proposal.} The occurrence of quasicrystalline or generally fluctuating IGF-induced NHSE and topological phase transition can be tested in tunable non-Hermitian systems. Examples include photonic quantum walks \cite{Lin2022,Weidemann2022},  acoustic crystal \cite{Wang2024}, and arrays of semiconductor lasers \cite{Gao2023}, where engineering of nonreciprocal hopping and bond disorder is viable. In a quasiperiodic laser array, when the system operates in proximity to the critical point at around $\lambda\sim 2$, it has the capability to support extended lasing modes that are pivotal for applications necessitating high-power emission over extensive areas.. Nonetheless, this requires an examination of non-equilibrium dynamics through the utilization of rate equations \cite{Longhi2022,Zhu2022}, a topic that exceeds the scope of the current investigation.

{\it Conclusion.} We have demonstrated a theoretical understanding of point-gap spectral topology, localization and transport characteristics induced by fluctuating IGFs in a non-Hermitian lattice gauge theoretical framework subject to broken translational symmetry. This is facilitated by a comparative analysis with imaginary gauge-transformed mean-field periodic lattices exhibiting equivalent spectral characteristics.  A hidden topological phase transition has been uncovered in the special case of a quasicrystalline IGF lattice. Our findings open possibilities for the design of tunable topological transport in devices that can accommodate flexibility with respect to periodicity and translation symmetry constraints.  The expansion of the theory to encompass lattices featuring non-Hermitian disorder in the onsite potential remains a pending challenge.

{\it Acknowledgments.}  { The research is supported by the Science and Engineering Research Board (SERB), India (Grant No.~MTR/2023/000249) and a Seed Grant from IISER Berhampur.}

\onecolumngrid

\setcounter{figure}{0}
\renewcommand{\thefigure}{S\arabic{figure}}

\setcounter{equation}{0}
\renewcommand{\theequation}{S\arabic{equation}}

\bigskip
\medskip

	\begin{center}
		{\large \bf Supplementary Material}
	\end{center}
	\medskip
	
{\bf Imaginary gauge field (IGF) induced long-range hopping.} We consider particle hopping in a one-dimension lattice under imaginary vector potential ${\bf A}=i{g(x)} \hat{x}$, where $g(x)=g_n \in \mathbf{R}$ for $n\le x<n+1$, is in general site dependent. The forward hopping amplitude from $n$th site to $(n+p)$th site is obtained by Peierl's substitution 
\begin{equation}
	J^{(p)}\mapsto J^{(p)} e^{i\theta_n^{(p)}},
\end{equation}
where $J^{(p)}$ is the Hermitian hopping coefficient to $p$th nearest-neighbor and $\theta_n^{(p)}$ is the site-dependent imaginary phase given by
\begin{equation}
 \theta_n^{(p)}=\int_{n}^{n+p} A\cdot dr= {i\left(\int_n^{n+1}+\int_{n+1}^{n+2} \cdots +\int_{n+p-1}^{n+p}\right) g(x) dx}= {i\left(\sum\limits_{j=n}^{n+p-1} g_j\right)}.
\end{equation}	
The amplitude of backward hopping, i.e., from $(n+p)$th to $n$th site is similarly obtained as $J^{(p)} e^{-i\theta_n^{(p)}}$. 

\begin{figure}[t!]
	\centering{
		\includegraphics[width=0.7\textwidth]{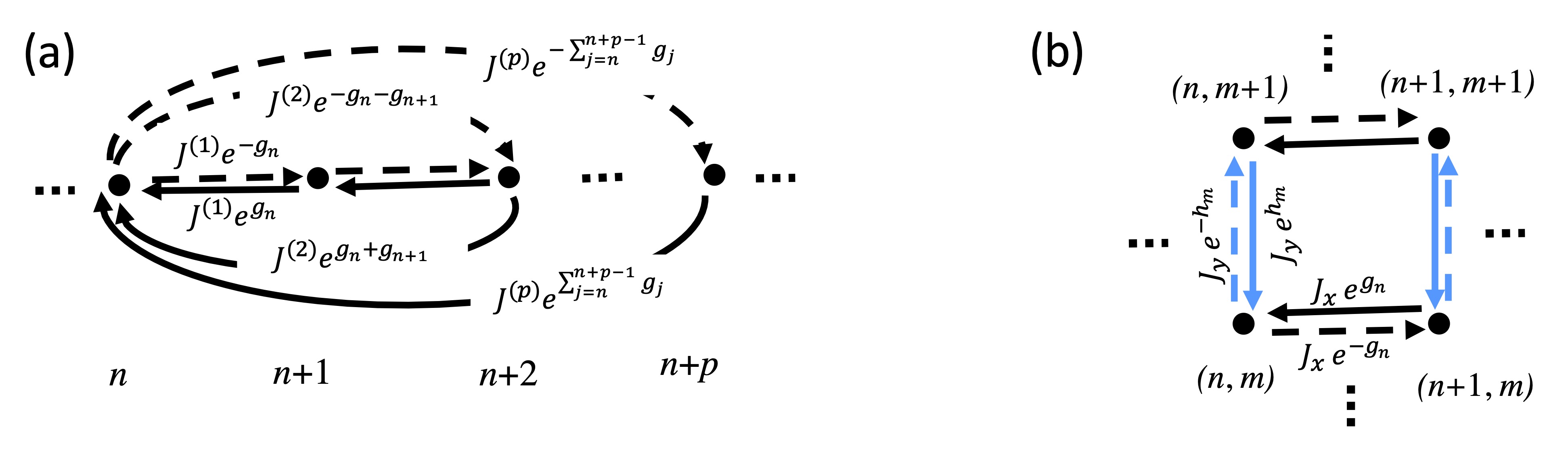} }
	\caption{Schematic of 1D and 2D lattices with IGF induced hopping. } \label{fig-s1}
\end{figure}

{\bf Extension of Hatano-Nelson model with IGF induced long range hoppings.} A  Hamiltonian, with open boundaries, which includes long range hopping terms can be written as  (Fig. S1a)
\begin{equation}
H=	\sum_{p=1}^P H^{(p)},\label{S3}
\end{equation}
where
\begin{equation}
	H^{(1)}=J^{(1)}  \sum_{n=1}^{N-1} \left(e^{g_n} c^+_n c^-_{n+1} +  e^{-g_n} c^+_{n+1} c^-_{n}\right)
\end{equation}
represents the nearest-neighbor interaction, and next-nearest-neighbor interaction is described by
\begin{equation}
	H^{(2)}=J^{(2)}  \sum_{n=1}^{N-2} \left(e^{g_n+g_{n+1}} c^+_n c^-_{n+2} +  e^{-g_n-g_{n+1}} c^+_{n+2} c^-_{n}\right).
\end{equation}
In general, Hamiltonian that accounts $p$-th nearest neighbor hopping is given by 
\begin{equation}
	H^{(p)}= J^{(p)}\sum_{n=1}^{N-p} \left( e^{-i\theta_n^{(p)}} c^+_n c^-_{n+p} +  e^{i\theta_n^{(p)}} c^+_{n+p} c^-_{n}\right)=J^{(p)}\sum_{n=1}^{N-p} \left( e^{g_n+g_{n+1}\cdots+g_{n+p-1}} c^+_n c^-_{n+p} +  e^{-g_n-g_{n+1}\cdots-g_{n+p-1}} c^+_{n+p} c^-_{n}\right).
\end{equation}
Here, all the Hermitian hoppings (in the absence of gauge-field), $J^{(p)}, p=1,2,3\cdots$ are assumed to be real valued and uniform through out the lattice.  For a lattice with closed boundary, the Hamiltonian is given by $H=\sum\limits_{p=1}^{P} (H^{(p)} + H_B^{(p)})$, where $H_B^{(p)}$ is the boundary term
\begin{equation}
	H_B^{(p)}=J^{(p)}\sum_{n=N-p+1}^{N}\left(e^{-i\theta_n^{(p)}} c_n^+c_{n+p}^- + e^{i\theta_n^{(p)}} c_{n+p}^+c_{n}^-\right),
	\end{equation}
 subject to the assumption $N+p\equiv p$.

 {\bf Imaginary gauge transformation (IGT) and isospectral Hamiltonians.} Under the IGT \eqref{IGT1} (see main text), the open boundary lattice Hamiltonian \eqref{S3} reduces to its isospectral Hermitian Hamiltonian 
 \begin{equation}
 	\tilde{H}=\sum\limits_{p=1}^P\sum\limits_{n=1}^{N-p} J^{(p)}  \left(\tilde{c}_n^+ \tilde{c}_{n+p}^-+\tilde{c}_{n+p}^+ \tilde{c}_{n}^-\right)
 	\end{equation}
 	which is free from both IGF and non-Hermiticity. Here we have used the following identity 
 	\begin{equation}
 		i\theta_n^{(p)}-G_n +G_{n+p}=\left(-\sum\limits_{j=n}^{n+p-1} - \sum\limits_{j=1}^{n-1}  +\sum\limits_{j=1}^{n+p-1}\right) g_j = 0
 		\end{equation}
 	The Hamiltonian $H$ and $\tilde{H}$ share entirely real spectra, but NHSE occurs in $H$ only. On the other hand, a closed boundary lattice Hamiltonian reduces to the following mean-field long-range Hamiltonian by the application of IGT \eqref{IGT2} (of main text):
 	\begin{equation}
 		\tilde{\tilde{H}}= \sum_{p=1}^{P}\sum_{n=1}^{N} J^{(p)}\left(e^{p\bar{g}}~\tilde{\tilde{c}}^+_n \tilde{\tilde{c}}^-_{n+p} + e^{-p \bar{g}}~ \tilde{\tilde{c}}^+_{n+p} \tilde{\tilde{c}}^-_{n}\right),
 	\end{equation}
  which can be further Fourier transformed to obtain the mean-field band energy $E(k)=\sum\limits_{p=1}^P 2 J^{(p)} \cos p(k-i\bar{g})$. Fig. S2 shows the open and closed boundary spectra and eigenstates for a lattice with nearest and next-nearest neighbor hoppings under quasiperiodic $g_n$  given in the main text (Eq. (12)).  The emergence of open boundary NHSE depends on the value of $\lambda$ and corresponding mean-field $\bar{g}(\lambda)$ (similar to nearest-neighbor model discussed in the main text). A notable feature of the higher-order hopping model is the  higher order winding of the closed boundary spectrum w.r.to some open boundary spectral points $E_j$ (see Fig.S2). 
  
   \begin{figure}[t!]
  	\centering{
  		\includegraphics[width=0.7\textwidth]{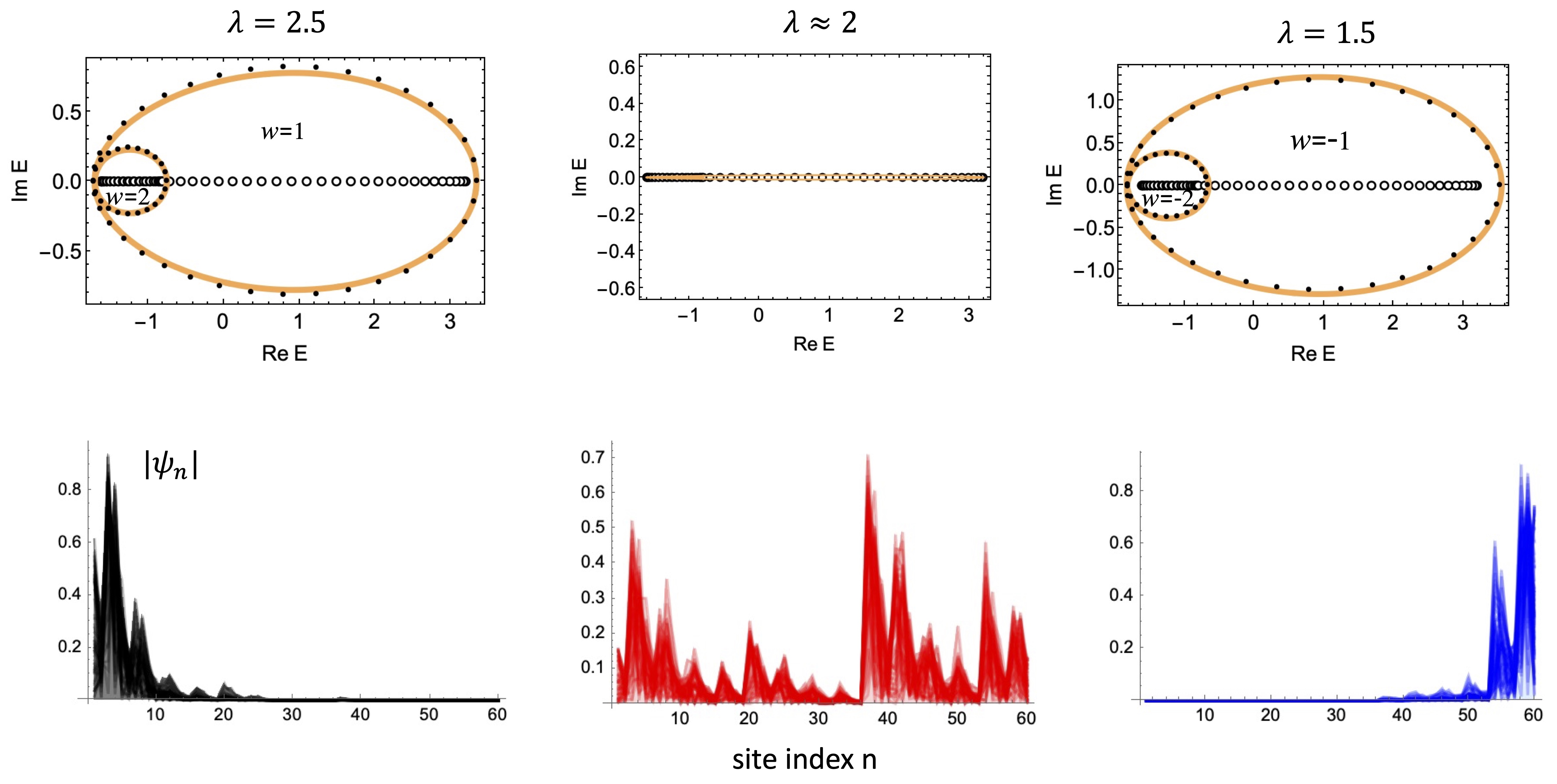} }
  	\caption{Spectra and eigenstates of a lattice under quasiperiodic IGF $g_n=\ln \lambda|\cos 2\pi \alpha n|$, $J^{(1)}=1, J^{(2)}=0.6 J^{(1)}$, and $N=60$. Open circles, black dots and orange line represent numerically obtained open boundary spectrum, closed boundary spectrum and mean-field spectrum, respectively. Left panel: $\lambda=2.5$, $\bar{g}=0.2$, Middle panel: $\lambda=2.06$, $\bar{g}=0.003$,  Right panel: $\lambda=1.5$, $\bar{g}=-0.3$.} \label{fig-s2}
  \end{figure}

 {\bf 2D Hatano-Nelson lattice under fluctuating IGF.} A lattice Hamiltonian $H_{2D}\{g_n,h_m\}$ (given in the main text with nearest-neighbor hopping, see also Fig.S1(b)) under 2D IGF $(ig_n\hat{x}+ih_m\hat{y})$ transforms into a spatially uniform lattice 
 \begin{equation}
 \tilde{H}_{2D}\{0,0\}=\sum_{n=1}^{N_x-1}\sum_{m=1}^{N_y-1} J_x\tilde{c}_{n,m}^+\tilde{c}_{n+1,m}^-+J_y\tilde{c}_{n,m}^+\tilde{c}_{n,m+1}^-+ h.c.
 \end{equation}
  under the 2D generalized gauge transformation (valid for open boundary condition)
 \begin{equation}
 	c_{n,m}^\pm= e^{\pm G_{n,m}}\tilde{c}_{n,m}^\pm
 	\end{equation}
 	where $G_{n,m}=\sum_{j=1}^{n-1}g_j+\sum_{\ell=1}^{m-1}h_\ell$. For a lattice with closed boundary in both $x$ and $y$ directions, one obtains mean-field Hamiltonian $\widetilde{\widetilde{H}}_{2D}\{\bar{g},\bar{h}\}\}$ by the IGT
 	\begin{equation}
 	c_{n,m}^{\pm}=e^{\pm G_{n,m}\mp (n-1)\bar{g}\mp (m-1)\bar{h}}\tilde{\tilde{c}}_{n,m}^\pm,
 		\end{equation}
  where $\bar{g}=\sum g_n/N_x$ and $\bar{h}=\sum h_m/N_y$ and $N_x, N_y$ are the numbers of sites along $x$ and $y$ directions, respectively. The energy dispersion of $\widetilde{\widetilde{H}}$ is obtained as $E(k_x,k_y) =2J_x \cos (k_x-i\bar{g}) +2J_y \cos(k_y-i\bar{h})$. Considering the separability of the Hamiltonian $H_{2D}=H_x~\bigoplus~H_y$, the weak topological invariance  (see Refs.~\cite{Gao2023},\cite{Kawabata2020} of main text) ${\bf{w}}=(w^x,w^y)$  of the system can be obtained from 1D winding numbers discussed in the main text. Corner NHSE occurs when both $w^x$ and $w^y$ are non-zero, whereas edge NHSE occurs when one of them is either zero or undefined depending on the mean-fields $(\bar{g},\bar{h})$. Fig.S3 demonstrates various situations for quasiperiodic IGF along $x$ and $y$ directions given by
  \begin{equation}
  	g_n=\ln |\lambda_x\cos 2\pi\alpha n|, \quad h_m=\ln|\lambda_y\cos 2\pi \alpha m|,
  	\end{equation} 
  respectively. The winding number  ${\bf{w}}(E): (1,1), (-1,-1),(1,-1)$ or $(-1,1)$ depends on $(\lambda_x,\lambda_y): (>2,>2), (<2,<2), (>2,<2)$ or $(<2,>2)$, respectively. Consequently, the eigenstate corresponding to $E$ localizes at the lower-left, upper-right, upper-left or lower-right corner, respectively. 
 
   \begin{figure}[t!]
 	\centering{
 		\includegraphics[width=0.6\textwidth]{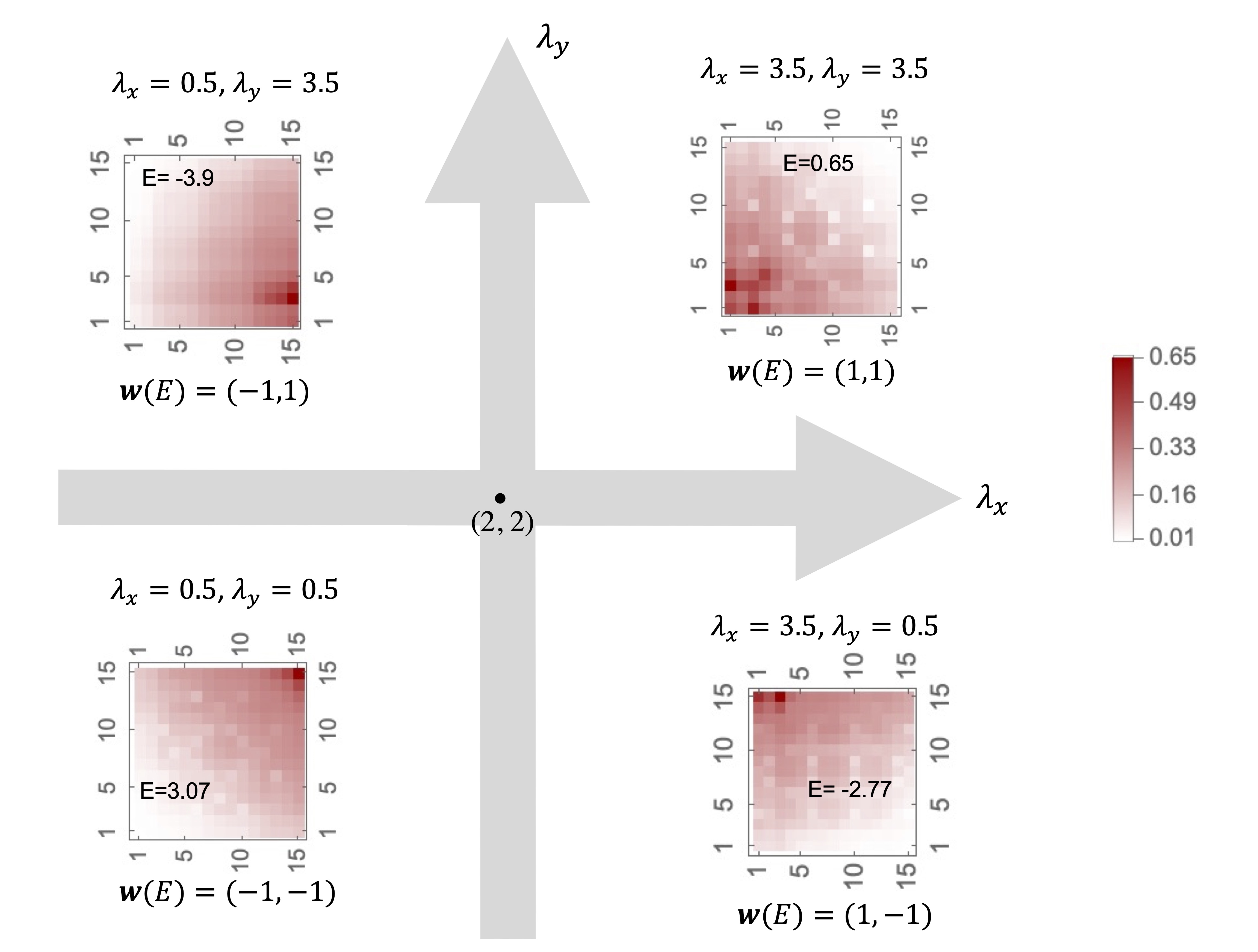} }
 	\caption{Examples of corner localization in a 2D open lattice with $(15\times15)$ sites. Absolute values of eigenstates are shown in for different choices of $(\lambda_x,\lambda_y)$. Here, $J_x=J_y=1$. } \label{fig-s3}
 \end{figure}
 
 {\bf Determination of mean-field approximation Eq.\eqref{mean-field}.} The approximation in Eq.~\eqref{mean-field} was obtained by numerical experiment. The figure below illustrates that the value of $\left(\prod_{n=1}^{N}|\cos2\pi\alpha n|\right)^{1/N}$ converges to $1/2$ for large $N$, hence the approximation follows.  
 
  \begin{figure}[hb!]
 	\centering{
 		\includegraphics[width=0.5\textwidth]{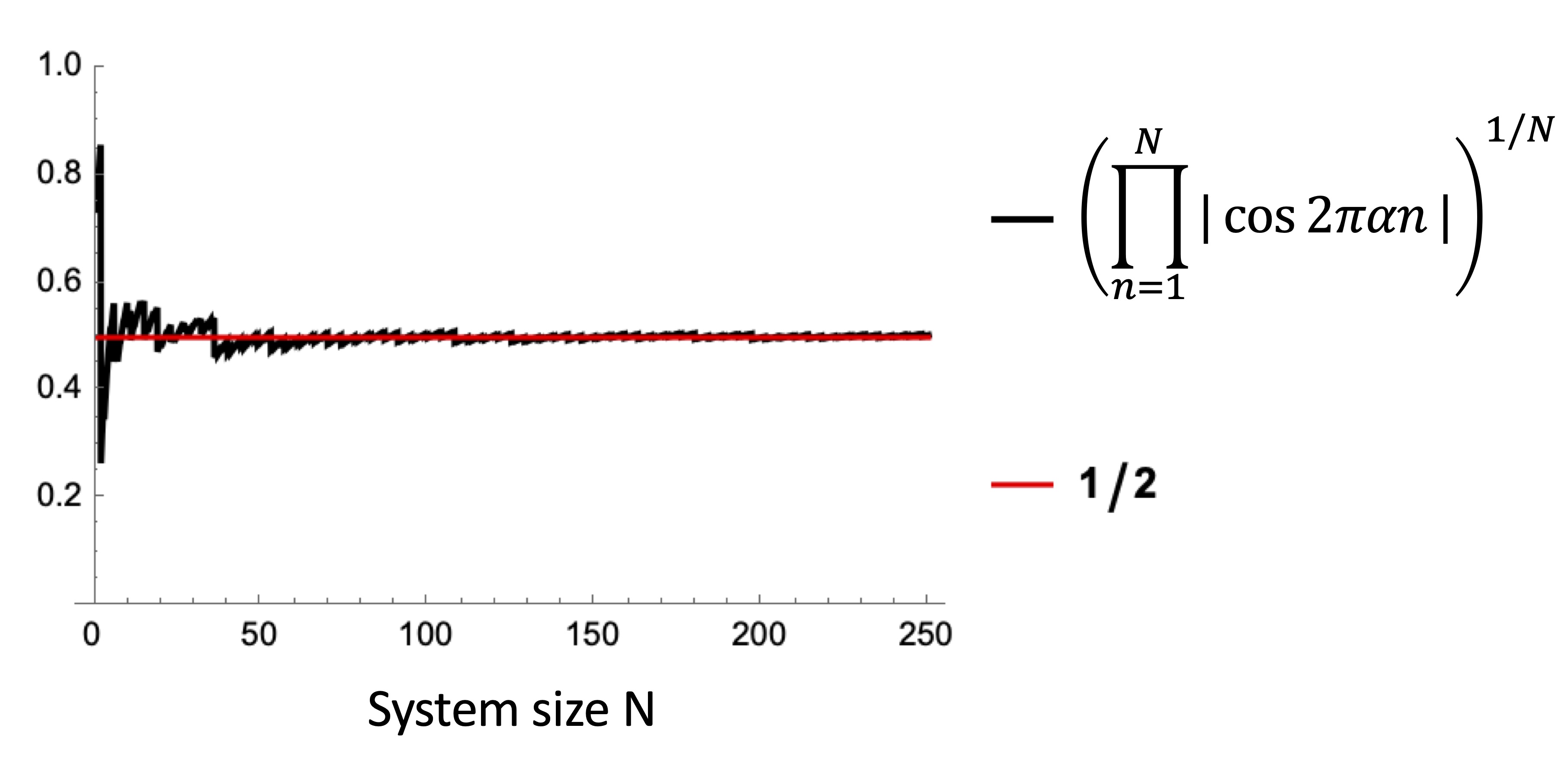} }
 	\caption{The geometric mean of $|\cos 2\pi\alpha n|$, $n=1,2,\cdots N$ converges to $1/2$ for large $N$, where $\alpha=(\sqrt{5}-1)/2$.} \label{fig-s4}
 \end{figure}
 
\twocolumngrid

	\end{document}